# Vocal effort modulates the motor planning of short speech structures

Alan Taitz[1], Diego E. Shalom[2] and Marcos A. Trevisan[1,2]

[1] Physics Institute of Buenos Aires (IFIBA) CONICET, Buenos Aires, Argentina,

[2] Department of Physics, Universidad de Buenos Aires, Buenos Aires 1428EGA, Argentina.

Corresponding author:

marcos@df.uba.ar (MAT)

March 2, 2018

**Speech requires programming the sequence of vocal gestures that produce the sounds of words. Here we explored the timing of this program by asking our participants to pronounce, as quickly as possible, a sequence of consonant-consonant-vowel (CCV) structures appearing on screen. We measured the delay between visual presentation and voice onset. In the case of plosive consonants, produced by sharp and well defined movements of the vocal tract, we found that delays are positively correlated with the duration of the transition between consonants. We then used a battery of statistical tests and mathematical vocal models to show that delays reflect the motor planning of CCVs and transitions are proxy indicators of the vocal effort needed to produce them. These results support that the effort required to produce the sequence of movements of a vocal gesture modulates the onset of the motor plan.**

## I. INTRODUCTION

Voluntary movements need preparation before execution [1]. This preparatory phase received extensive attention from the scientific community, whose investigations across species [2,3] helped unveiling the instructions to the effectors before tasks as locomotion [4], arm reaching [5] and vocal production [6,7]. In humans, the readiness potential (RP) is considered a universal signature of planning of volitional acts that has been largely investigated with electroencephalographic techniques [8–10].

Speech requires programming the instructions to drive two acoustically uncoupled effectors: the vocal folds and vocal tract [11]. The folds are a pair of opposed membranes located at the larynx, which leads to a series of configurable cavities ending at the mouth, known as vocal tract. When the larynx is adducted and lung pressure reaches a threshold, the folds oscillate colliding with each other [12]. The resulting periodic interruptions of airflow produce a pressure wave that propagates along the vocal tract. When the tract is open, the sound radiated from the mouth corresponds to a vowel [13]. Consonants are produced in different manners: fricatives are vortex sounds produced by the turbulent passage of air through a constriction in the vocal tract [14]. When lips are narrowed, an [f] is produced; when the tip of the tongue approaches the upper teeth, we hear [s]. Plosives are generated by completely occluding the vocal tract at a given point [15]. When lips are released, we hear [b]; releasing the tip or the body of the tongue produces a [d] or a [g], respectively. From a mechanical point of view, these families of sounds represent vocal gestures of different effort: vowels are produced by smooth and opened tract configurations, fricatives by narrow constrictions and plosives by sharp closures.

Recent advances give a cause for great optimism in describing vocal motor control in the language of dynamical systems [5]. As a matter of fact, the avian vocal system presents strong analogies with the human case [16], and birdsong has been fully described in dynamical terms [7,17], leading to devices that generate song controlled by a few physiological variables [18].
What are the dynamical variables driving the human vocal system? The production of speech requires coordination of the larynx, lips, tongue and jaw to produce a gesture, and then control of 'the strength with which the associated gesture (e.g., lip closure) "attempts" to shape vocal tract movements at a given point in time' [19]. These two variables, the gesture and its strength, have been interpreted in dynamical terms [19]. In this theoretical description, the vocal program is executed by coupled neural oscillators; their relative phases represent the sequence of movements of the articulators that define a gesture, and their amplitude represents its strength.

Here we hypothesized that the vocal program is sensitive to the strength of the vocal gestures, and that this dependence is revealed in the timing between the instruction and the vocal onset. To explore this hypothesis, we created and analyzed a database of speech structures produced under a simple reaction-time paradigm.

## II. MATERIALS AND METHODS

A. Participants

Forty five native Spanish speakers (23 females, age range 20-28, mean age 22), undergraduate and graduate students at the University of Buenos Aires, with normal or corrected-to-normal vision and no speech impairments, completed the experiment. All the participants signed a written consent form. All the experiments described in this paper were approved by the ethics committee Comité de Ética del Centro de Educación Médica e Investigaciones Clínicas 'Norberto Quirno' (CEMIC) qualified by the Department of Health and Human Services (HHS, USA): IRb00001745-IORG 0001315.

B. Stimuli and tasks

Participants sat in a silent room, 0.7 m away from a 19' monitor. They were asked to pronounce a sequence of consonant-consonant-vowel (CCV) structures, as soon as they appeared at the center of the screen. The CCVs were recorded with a commercial microphone at approximately 0.2 m of the mouth.

After pronouncing a CCV, the participant was instructed to press a key. A blank screen preceded the following CCV, which appeared after a random time between 0.5 and 2 s. The procedure was repeated until completing 3 rounds of the complete CCV set. All the presentations were randomized.

The experiment was coded in Python, using Pyaudio and PsychoPy [20] libraries to ensure robust audio-video synchronization (30 ms maximum offset between visual presentation and audio recording).

CCVs were formed by combinations of fricatives and plosives, followed by the vowel [a]. We used the most common Spanish fricatives [21,22] [f, s, x], pronounced as in **f**ace, **s**tand and lo**ch**, and the most common Spanish plosives [b, p, d, t, g, k] [21], pronounced as in **b**ay, **p**ay, **d**ie, **t**ie, **g**ray and **c**ray. Plosives come in voiced-unvoiced pairs [b, p], [d, t] and [g, k]. The tract anatomy is identical for each pair, but folds are active only for the first components. Since plosives are recognized through the sound produced at the occlusion release, unvoiced ones leave no acoustical traces during occlusion. On the contrary, the folds' fundamental frequency serves to spectrally mark vocal onset during occlusion in voiced plosives. For this reason, CCVs starting with plosives were reduced to the voiced ones [b, d, g].

We presented the 6 possible combinations of fricatives (*fja, fsa, jfa, jsa, sfa, sja*), the 9 possible plosive-fricative combinations (*bfa, bja, bsa, dfa, dja, dsa, gfa, gja, gsa*), a selection of 11 fricative-plosive combinations (*fga, fba, fpa, jba, jca, jga, jda, sba, sca, sga, sta*) and 11 combinations of plosives (*bca, bda, bga, bta, dba, dca, dga, dta, gba, gda, gta*). This set of 37 CCVs allowed keeping a short (mean 20 minutes), single-session experiment per participant.

The number of recorded samples was 4995 = 45 subjects × 3 trials × (6 fricative-fricative + 9 plosive-fricative +11 fricative-plosive + 11 plosive-plosive) combinations.

C. Data selection

Stage 1.

Participants were instructed to pronounce the CCVs clearly, as soon as possible after visual presentation, and they were not allowed to correct their utterances. These requirements produced a considerable amount of errors during the experiment. We discarded audio files that a. contained more than one attempt to produce the CCV and b. did not match with the presented CCV. This left us with 3296 CCVs.

Stage 2.

For each audio file that passed stage 1, we measured three timing quantities: the delay $\Delta$ between visual presentation and voice onset, the transition $\tau$ between consonants and duration $T$ of the CCV. Audio files were inspected with Praat [23].

Delays $\Delta$ were measured directly from the spectrogram. Transitions are defined as the time $\tau$ during which the vocal tract changes its configuration from the first to the second consonant. Since vocal gestures producing the same speech content are not necessarily unique [24], we extracted by inspection the most robust features of each CCV across trials and participants, and therefore only selected the samples that strictly met the following spectral signatures for consonantal transitions:

a. *Fricative-fricative combinations*. As noisy turbulent sounds, the spectral signature of a fricative is a dark spot, such as the ones shown in the left panel of Fig. 1A for [f] and [s]. Transitions are defined by the continuous passage from the first to the second constriction, therefore presenting a mixture of spectral components from both consonants.
b. *Plosive-fricative combinations*. Transitions are characterized by the interval that goes from releasing the occlusion of the plosive to the formation of a constriction to generate the fricative. Spectrally, the transition is described by the voiced sound structure produced after the plosive, and before the formation of the purely noisy fricative spot (Fig. 1B, left).
c. *Fricative-plosive combinations*. During these transitions, the vocal tract evolves from a constriction at one point to an occlusion at another one. Transitions are spectrally defined as the interval between the abrupt end of the fricative and the release of the plosive into the vowel [a], as shown in the left panel of Fig. 1C.

d. *Plosive-plosive combinations.* Vocal tract passes from an occlusion at a given point of the tract to another one at a different location. Transitions, as shown in the representative case of the left panel of Fig. 1D, go from the release of the first plosive characterized by a voiced sound structure, to the release of the second one within the vowel [a].

A pool of 1958 CCVs was successfully classified into these spectral categories. Four subjects were discarded for presenting systematical spectral differences from the described categories.

Stage 3.

We finally discarded data that passed stage 2 for which delays $\Delta$ and transitions $\tau'=\tau/T$ were greater than two standard deviations from the speaker mean values. A final timing dataset from 1519 CCVs from N=41 participants was used to perform the analyses presented in this work.

D. CCV frequency

None of the CCVs are Spanish words. We counted the intra-word appearances of each CCV in a large Spanish corpus [25], and assigned the appearances per million words as the frequency of each CCV. Since the range of frequency values was large (from 0 to roughly $10^5$ apmw), in Fig. 2 we show a discretization of this range in low, medium or high frequency values.

E. Vocal model

The main ingredients of the folds' dynamics are caught by the following equation of motion, known as the flapping model [26]:

$$m\ddot{z} = -k(z)z - b(z,\dot{z})\dot{z} + p_s \frac{\delta + 2\Gamma\dot{z}}{z_0 + \delta/2 + z + \Gamma\dot{z}} \quad (1)$$

In Fig. 3A we sketch its elements. The variable $z$ represents the transverse displacement of the folds from the pre-phonatory position $z_0$. The tissue is described by nonlinear elastic $k$ and dissipative $b$ functions. The last term represents the energy transferred to the folds by lung pressure $p_s$, which depends on the parameters $\delta$ and $\Gamma$ that define the folds' configuration. Dynamical analysis of Eq. 1 revealed that pressure $p_s$ beyond a threshold creates oscillations through a Hopf bifurcation [12] and that the threshold increases with glottal abduction $z_0$ [27]. Oscillations make the folds collide, interrupting the airflow at the vocal tract entrance and producing pressure waves $p$ that propagate along a vocal tract. The tract cross-section $A(x,t)$ is shown in Fig. 3B, where $x$ represents the distance from the glottis ($x=0$) to the mouth ($x=L$). During a general sequence of vowels and plosive consonants, $A(x,t)$ was described by Story [28]:

$$A(x,t)=\pi/4[\Omega(x)+q_1(t)\phi_1(x)+q_2(t)\phi_2(x)]^2 \prod_{k=1}^{n}[1-c_k(x)m(t-t_k)] \qquad (2)$$

The first factor represents the vowel substrate, with empirical functions $\Omega$, $\varphi_1$ and $\varphi_2$ obtained from an orthogonal decomposition calculated from MRI data [29]. Any sequence of vowels is generated by the evolution of $q_1$ and $q_2$. Plosive consonants are generated through the bell-shaped functions $c_k$ and $m$. These functions reach the value 1 around $x=x_k$ and $t=0$ respectively, producing a specific occlusion $A(x_k,t_k)=0$. The spatial functions $c_k$ represent the anatomy of the occlusion for each consonant, and the time function $m$ represents its activation-deactivation.

F. CCV synthesis

Eqs. 1 and 2 were solved with MATLAB. Numerical integration of Eq. 2 was performed through a standard Runge-Kutta algorithm at a sampling rate of 44.1 kHz. A wave-reflection model [30,31] was used to simulate propagation of sound along the vocal tract, which is approximated by a series of $N=44$ tubes of cross-section $A_i$, $1 \leq i \leq N$. Sound propagation is solved by splitting an incoming sound wave $p_i$ into a reflected and a transmitted wave at each interface, with reflection and transmission coefficients $r_{i,j}=(A_i-A_j)/(A_i+A_j)$ and $t_{i,j}=1-r_{i,j}$ ($j=i\pm1$) [32]. The sound radiated from the mouth is proportional to the pressure at the last tube, $p_{44}$. The resulting time series was converted to wav audio format. To synthesize CCVs for voiced plosives, we proceeded as follows:

1. Subglottal pressure $p_s$ was set at a normal speech value such that Eq. 1 generated oscillations of the folds with a pitch around 100 Hz, from $t=0$ to voice offset at $t=T_f$. The rest of the parameters can be found elsewhere [11].
2. Linear functions $q_1(t)=4t/T$ and $q_2(t)=t/T$ [26] were used in Eq. 2 to drive the tract from a neutral tract $(q_1,q_2)=(0,0)$ at $t=0$ to the shape of the vowel [a] $(q_1,q_2)=(4,1)$ at $t=T$, and maintained there until $t=T_f$.
3. While the tract evolves towards the vowel [a], two occlusions take place at $t_1$ and $t_2$, activated through Gaussian functions $m$ of width 0.14 s [28], separated by $\tau=t_2-t_1$. We used the mean value of τ across CCVs and participants. The spatial functions $c_k$ reach the value 1 at $x_k$~$L$ for [b], $x_k$~$9/10L$ for [d] and $x_k$~$7/10L$ for [g]. Functional forms and parameter values for the functions $c_k$ and $m$ are reported elsewhere [28,33].

## III. RESULTS

We investigated the vocal planning prior to the utterance of short speech structures. For that sake, we recorded 45 speakers while pronouncing a set of consonant-consonant-vowel structures (CCV) as soon as they appeared on a computer screen. We used combinations of fricatives [f, s, x] and plosives [b, p, d, t, g, k] followed by the vowel [a]. Examples of these structures are *fsa*, *gfa*, *fga* and *dca*, which are also representatives of fricative-fricative, plosive-fricative, fricative-plosive and plosive-plosive combinations respectively.

For each audio file, we measured three timing variables: the delay between visual presentation and phonation onset (Δ), the duration of the CCV (*T*) and the duration of the transition between the consonants (τ). In the left panels of Fig. 1 we show these timing variables for representative cases of the fricative-fricative (A), fricative-plosive (B), plosive-fricative (C) and plosive-plosive (D) types. After selecting the audio files that met the spectral criteria for each family type (see section II C), we collected timing data from 1519 CCVs produced by 41 subjects.

How are these timing variables related? To compare dynamics across participants with different speech rates, we normalized the transitions to the duration *T* of the CCV, τ'=τ/*T*. In the right panels of Fig. 1 we plot Δ vs. τ' for the different families of consonantal combinations. No correlations were found between delays and transitions for fricatives (Fig. 1A, *r*=-0.27, *p*=0.6) or combinations of fricatives and plosives (Fig. 1B, *r*=-0.29 *p*=0.38 and Fig. 1C, *r*=0.15, *p*=0.7). Interestingly, however, a positive correlation emerged for combinations of plosive consonants (Fig. 1D, *r*=0.68, *p*<0.02). We next explored how these timing variables relate to relevant motor and planning variables.

## A. Fricative-fricative combination

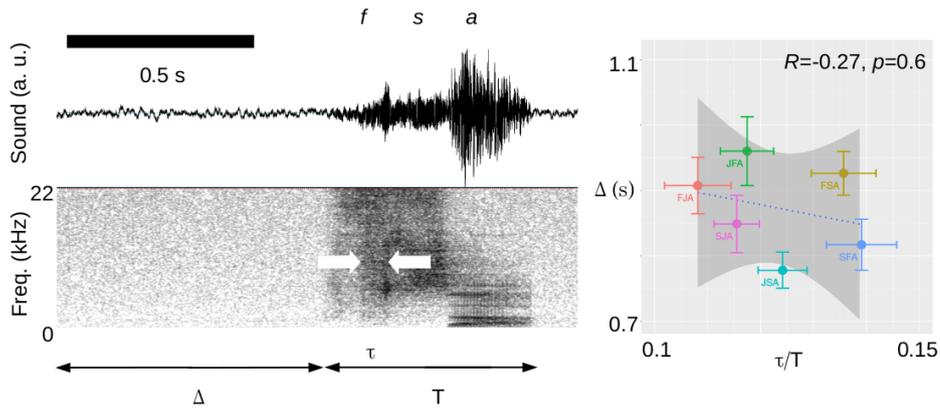

## B. Fricative-plosive combination

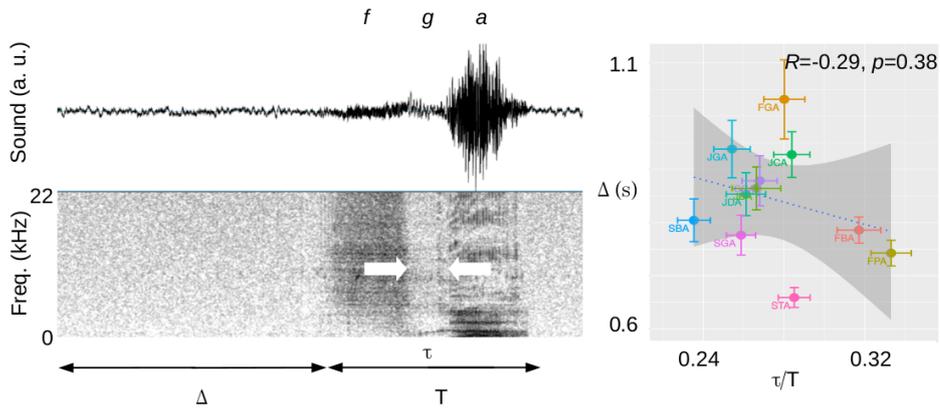

## C. Plosive-fricative combination

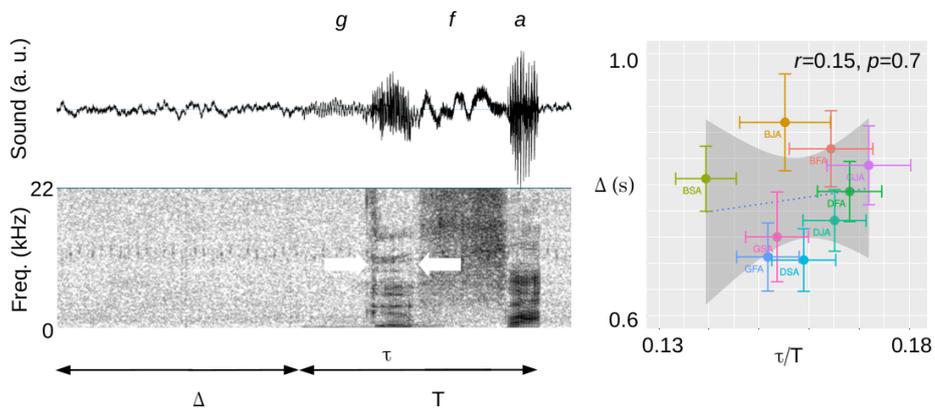

## D. Plosive-plosive combination

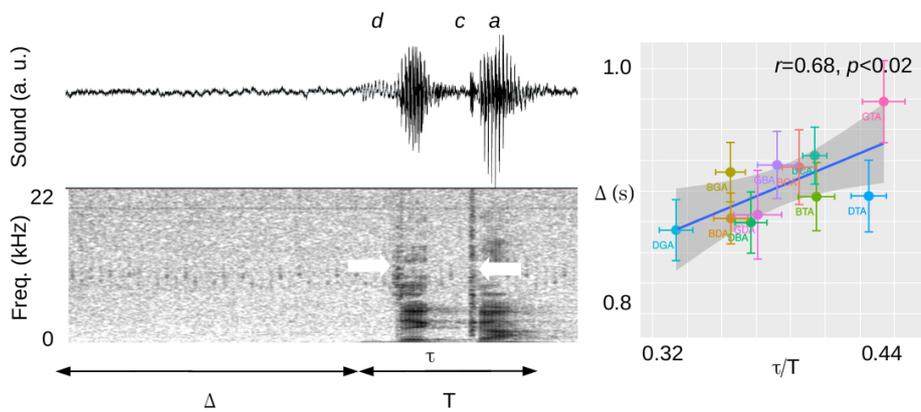

**FIG. 1. Timing variables for the different consonantal transitions.**

Left panels show onset delays Δ and consonantal transitions τ measured from sound and spectrogram, for representative cases of the fricative-fricative (A), fricative-plosive (B) plosive-fricative (C) and plosive-plosive (D) types. In the right panels we show that vocal onset delays Δ are not significantly correlated with the normalized transitions τ'=τ/T for the 6 fricative-fricative (A), 11 fricative-plosive (B) and 9 plosive-fricative (C) combinations. On the contrary, timing variables are positively correlated for the 11 plosive-plosive combinations (D).

A. Delays and motor planning

Vocal onset delays represent the integration of reading, memory retrieval and vocal planning tasks, followed by articulatory movements produced before phonation. We first analyzed the effects of each of these contributions.

Reading. Reading task involve visualization and recognition of three common Spanish phonemes of identical visual size. Given the short length of the strings, we assumed that this effect is constant across CCVs.

Memory retrieval. We explored two memory variables. Long term effects were investigated by computing the frequency of occurrence of the CCVs in Spanish, shown in Fig. 2 (see details in section IV D). Since our participants completed three rounds of the experiment, trial number was used to investigate short term memory effects. We submitted timing data to ANOVA with frequency and trial as independent variables (Table I). The analysis revealed effects of frequency on Δ and τ' for all the consonant combinations, except for the case of plosive-plosive type. This merely reflects that these are very unusual combinations in Spanish, and therefore no long term memory effects were detected. Trial number, on the contrary, has a significant effect on Δ and no effect on τ', for any consonantal combination. A follow up of this analysis revealed that this effect on Δ was accounted for by an increase in Δ for trial 1 (mean 0.94±0.01 s) than for trials 2 and 3 (mean 0.85±0.01 s), ($t$=8.13, $df$=1176, $p$<$10^{-14}$). We therefore considered trial 1 as a training set and used timing data from trials 2 and 3 for the following analyses.

Articulatory kinematics. Since Δ was measured from visual presentation to vocal onset, articulatory kinematics that normally precede phonation [34,35] are included within this time period (e.g. lip closure before folds' activation for an initial [b]). To study the effect of movements prior to sound onset, submitted our data to ANOVA using the first consonant of the CCV as independent variable, restricting the study to initial plosives [b], [d] and [g]. The production of these consonants before phonation require articulation of the lips, tongue tip and tongue body movements respectively. This analysis revealed no significant effects of initial plosives on Δ ($p=0.323$, $F=1.13$, $df=2$), which allowed us to assume that the period of articulation previous to phonation is constant across CCVs.

Due to the nature of the experiments performed here, vocal planning cannot be explored directly. Instead, we analyzed the different tasks performed by the participants from CCV visualization to phonation. We found that, for plosive-plosive combinations, the tasks of reading, memory and prephonatory movements can be assumed constant across CCVs. We therefore assumed that variations of Δ across CCVs account for differences in the vocal planning.

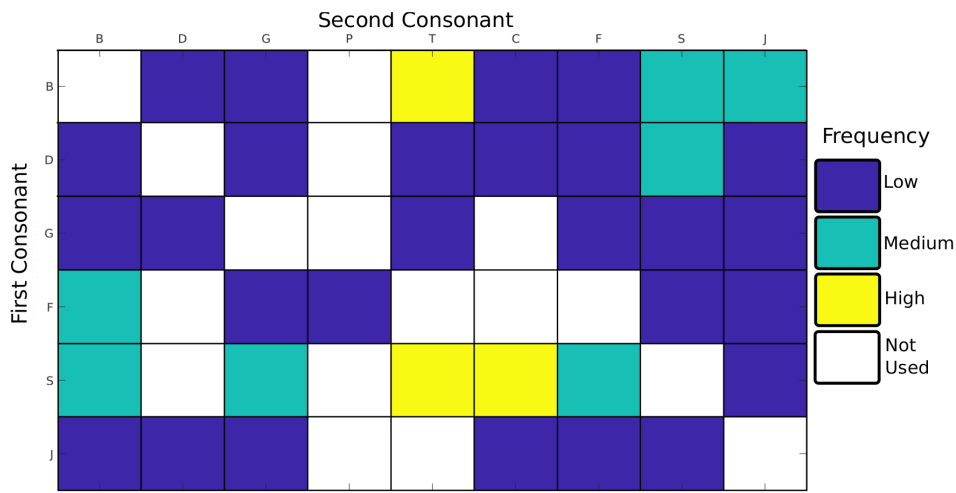

**FIG. 2. Frequency of occurrence of consonants in CCVs.**
Rows and columns correspond to the first and the second CCV consonants respectively. Colors represent low (blue), medium (light blue) and high (yellow) frequency of occurence values of the these combinations in Spanish. Occurrences of fricative-fricative and fricative-plosive combinations present high variability, while only one plosive-plosive combination presents a high occurrence level.

## B. Consonantal transitions and vocal effort

Under the simplistic hypothesis that effortless gestures are produced faster than harder ones, we could roughly assume that transitions τ' reflect the effort needed to produce a CCV.

We first note that, consistently with this hypothesis, the right panels Fig. 1 show that the shorter transitions correspond to fricatives (τ'=0.124±0.002), which require relatively low efforts to produce constrictions in the tract; longer transitions were found in the production of plosive-fricative (τ'=0.163±0.003) or fricative-plosive combinations (τ'=0.280±0.003), and these are in turn shorter than transitions between plosives (τ'=0.389±0.003), which require the largest efforts to completely occlude the vocal tract twice.

To further formalize this notion of effort, we propose a physically-based model that represents the effort $E$ exerted by the tract and the folds during speech. The expression reads:

$$E = E_{tract} + E_{folds} = \int_0^T \int_0^L |A(x,t) - A_\Omega(x)| dx\, dt + E_0 \qquad (3)$$

The shape of the vocal tract is mathematically described by its cross-section $A(x,t)$, with $x$ the distance from the glottis ($x=0$) to the mouth ($x=L$), as shown in the sketch of Fig. 3B. Building on a previous mass-spring model [27], the first term of Eq. 3 represents the elastic force exerted by the departures of the tract from a relaxed configuration $A_\Omega(x)$ along an utterance of duration $T$.

During a CCV of plosive consonants, two occlusions separated by τ occur while the tract is evolving towards the vowel. This is described in Eq. 2 (section II E), where plosives are spatially described by functions $c_k(x)$ that model the extension and location of the occlusion, and temporally described by functions $m(t)$ that control the activations of closures. Using the parameters reported for consonantal activations [19,28], efforts computed using each CCV transition (in a range that goes from $\tau_{DGA}$=0.33±0.03 s to $\tau_{GTA}$=0.45±0.04 s, shown in Fig. 3C) display the same organization as the efforts computed for all the CCVs using the population mean (τ=0.40±0.04 s) for all CCVs. To build confidence on the pertinence of these temporal parameters, we used Eqs. 1 and 2 to generate synthetic CCVs with the activations reported (see section II F) and fixed τ mean value, obtaining audio files with the same spectral features and speech content than the experimental records (Supplementary Materials).

Tract effort does not explicitly depend on τ, but instead on the specific consonants forming the CCV and their order of appearance. The dependence on the particular CCV consonants stem from the anatomical differences between them. For instance, the function $c_k(x)$ that models the extension of occlusion is wider for [d] than for [b], accounting for the greater extent where the constriction area is zero when it is performed with the tongue than with the lips; a greater effort is then needed to produce a [d] than a [b].

The dependence of the effort with the order of appearance of the consonants is due to the fact that consonants occur while the vocal tract is evolving towards the vowel [a]. For instance, *bda* requires less effort than *dba*. This asymmetry is explained because in *dba,* the lips closure for [b] occurs closer to the [a], for which the mouth is fully opened, requiring a larger vocal tract deformation that corresponds to a larger effort. Transitions $\tau'$ systematically reflect this asymmetry: $\tau'_{DGA} < \tau'_{GDA}$ ($t=-2.9$, $df=49.8$, $p<0.01$); $\tau'_{BGA} < \tau'_{GBA}$ ($t=-2.13$, $df=86.9$, $p<0.05$) and $\tau'_{BDA} < \tau'_{DBA}$ ($t=-0.71$, $df=136.5$, $p=0.48$).

The second term in Eq. 3 refers to laryngeal efforts. Vibration of the vocal folds has an on/off function in speech. Within our set of CCVs, vibration is either active during the whole vocalization (for the subset *bda, dba, bga, gba, dga, gda*) or inactivated during the second (unvoiced) plosive (subset *bca, bta, dca, gta, dta*). This second group involves a devoicing effort by glottal abduction. We model $E_{folds}$ in Eq. 3 as a constant that takes the value $E_0=0$ for the first group and $E_0>0$ for the second one. The distribution of CCVs along $E$ varies with $E_0$. For instance, a sufficiently large value of $E_0$ separates one group from the other, eventually producing a bimodal distribution. We set the value of $E_0$ that matched the experimental distribution of CCVs along $\tau'$ (vertical axis of Fig. 3C) to the theoretical distribution along $E$ (horizontal axis of Fig. 3C). We obtained values of $E_0$ between 7% and 11% of the mean vocal tract effort (for distributions between 4-8 bins).

The effort model of Eq. 3 represents a compromise between mathematical simplicity and vocal biomechanics, accounting for both tract reconfigurations and control of folds, which are the main gestures necessary to generate speech. Although effort does not depend on $\tau'$ within the range of measured values, Fig. 3C shows that both variables are significantly correlated ($R=0.84$, $p<0.001$), reflecting that consonantal transitions are indeed shorter when the global gesture requires less effort. This implies that $\tau'$ is a good proxy for the effort needed to produce the CCVs.

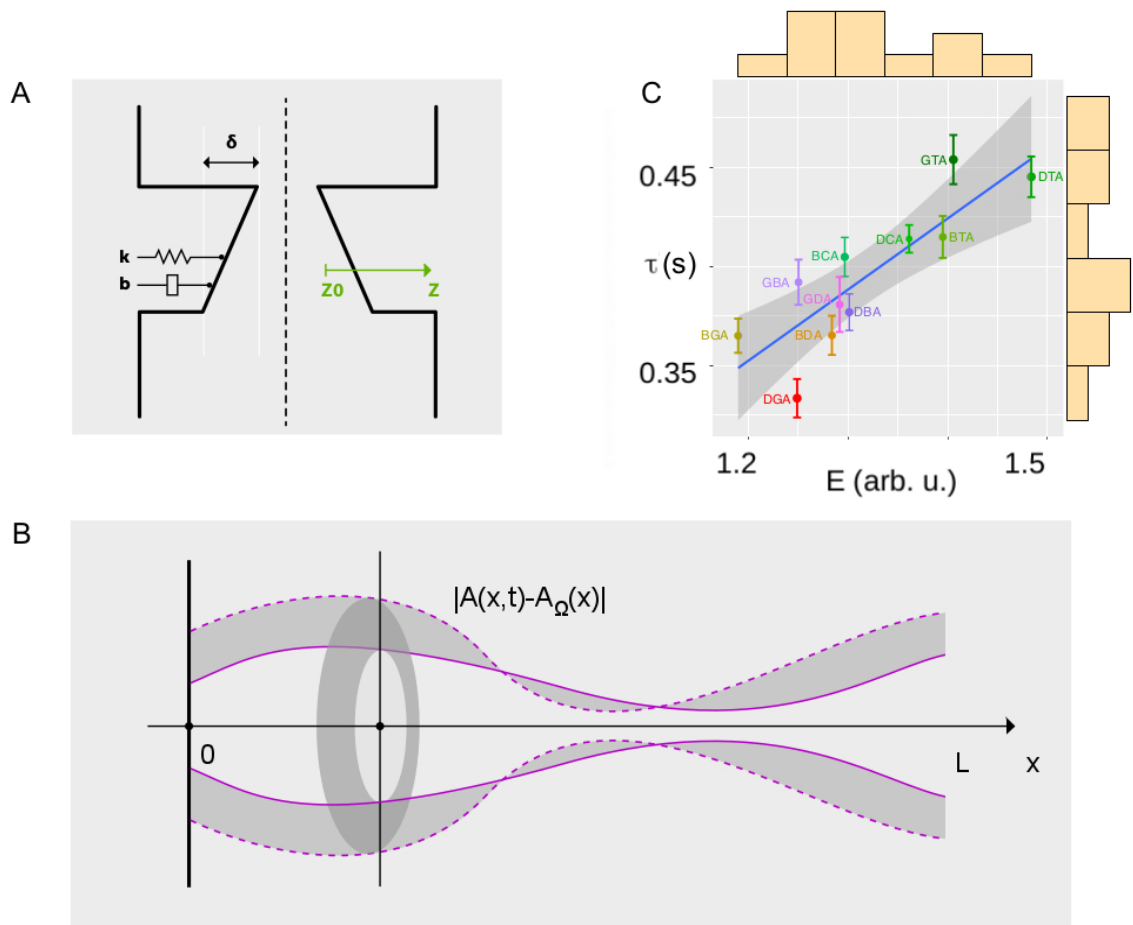

**FIG. 3. Elements of the articulatory model used to calculate vocal effort.**
A. Vocal tract movements, described by its cross-section $A(x,t)$, require an effort proportional to its deformations with respect to a relaxed configuration. B. Elements of the vocal folds model of Eq. 1. Abduction effort is needed to separate the folds, increasing $z_0$ in Eq. 1 to deactivate oscillations during unvoiced plosives. This is modeled as a constant $E_0$ for the group (*bca, bta, dca, gta, dta*) shown in green. C. The transitions $\tau$ measured on the recorded CCVs are positively correlated with vocal effort $E$ computed with Eq. 3. The distributions of CCVs with $E$ and $\tau$ are shown along the horizontal and vertical axis respectively, for $E_0$=9% of the mean tract effort.

Taken together, these results show that the timing variables measured directly from speech recordings are related to planning and motor variables. For combinations of plosive consonants, we found that differences in $\Delta$ account for differences in the motor planning and that $\tau'$ are proportional to the effort needed to produce the global CCV gesture.

## IV. CONCLUSIONS AND DISCUSSION

Here we hypothesized that the vocal program is sensitive to the strength of the vocal gestures, and that this dependence is revealed in the time required to activate vocal production. To explore this, we created and analyzed a database of speech structures produced under a simple reaction-time paradigm. Our participants were instructed to pronounce a set of consonant-consonant-vowel structures (CCV) as soon as they appeared on screen. We measured two timing quantities from the recorded audio files: the delay between visual presentation and vocal onset, and the duration of the transition between consonants, normalized to the duration of the CCV.

Using a battery of tools that includes statistical tests and nonlinear modeling of the vocal system, we found that 1. these variables are positively correlated for CCVs formed with plosive consonants. Then, we show the plausibility that 1. variations in delays account for variations in vocal planning and 2. consonantal transitions reflect the vocal effort required to produce the CCVs.
Overall, these results support the idea that the time required to plan short speech structures is proportional to the effort needed to produce them. The fact that no correlations were found for the rest of the consonantal combinations can be due to multiple factors. In particular, the memory effects detected on these combinations suggests a probable interaction between motor planning and memory tasks.

Our findings are indeed compatible with electroencephalographic measurements of the readiness potential (RP), defined as the buildup of electrical potential beginning about one second before cued or self-paced movement onset. This universal signature of planning and initiation of volitional acts [8] has been largely explored, and previous works reported that the level of force and speed of a motor task can be detected from these cortical potentials [10]. Moreover, differences in RP onset time account not only for initial movements but also for a sequence of movements [36].

Evidence is accumulating towards the vision of motor cortex as a dynamical system that generates and control movements [5]. The particular case of the vocal motor program has been extensively studied and successfully described in dynamical terms for songbirds [7]. Although strong analogies connect the avian and human vocal systems, descriptions of the human motor program in terms of dynamical systems received considerably less attention. One exception is the approach proposed by Goldstein and collaborators [19], whose theoretical model represents the vocal motor program as a set of coupled neural oscillators controlling the vocal articulators. The relative phases of the oscillators and their amplitudes configure a vocal gesture and its strength.

The model of vocal effort presented here presents a plausible quantification of this notion of vocal strength, supporting the hypothesis that the timing of the motor program is sensitive to the strength of the vocal gestures.

It is well established that brain motor areas activate not only for voice production, but also to assist other non-motor tasks as speech perception [37]. More interestingly perhaps, these motor maps are also used in the case of purely mental operations such the inner speech, i.e. the voice that "narrates the words we read or the conversations we imagine" [38]. In fact, as happens with speech itself, inner speech is associated with an efference copy with detailed auditory properties [38]. In light of these results, we plan to capitalize on the simple experimental paradigm presented here to explore vocal motor signatures during silent lecture.

## Acknowledgements

The research reported in this work was partially funded by the Concejo Nacional de Investigaciones Científicas y Técnicas, CONICET; The University of Buenos Aires UBA and NIH through R01-DC-012859.

## References

[1]   M. S. Graziano, Neuron **71**, 387 (2011).

[2]   W. A. Friedman, L. M. Jones, N. P. Cramer, E. E. Kwegyir-Afful, H. P. Zeigler, and A. Keller, J Neurophysiol **95**, 1274 (2006).

[3]   A. Riehle and J. Requin, J. Neurophysiol. **6**, (1989).

[4]   G. S. Stent, W. B. Kristan, W. O. Friesen, C. a Ort, M. Poon, and R. L. Calabrese, Science **200**, 1348 (1978).

[5]   M. M. Churchland, J. P. Cunningham, M. T. Kaufman, J. D. Foster, P. Nuyujukian, S. I. Ryu, and K. V Shenoy, Nature **487**, 51 (2012).

[6]   R. G. Alonso, M. a. Trevisan, A. Amador, F. Goller, and G. B. Mindlin, Front. Comput. Neurosci. **9**, 1 (2015).

[7]   A. Amador, Y. S. Perl, G. B. Mindlin, and D. Margoliash, Nature **495**, 59 (2013).

[8] A. Schurger, J. D. Sitt, and S. Dehaene, Proc. Natl. Acad. Sci. **109**, E2904 (2012).

[9] V. Siemionow, G. H. Yue, V. K. Ranganathan, J. Z. Liu, and V. Sahgal, Exp. Brain Res. **133**, 303 (2000).

[10] M. Jochumsen, I. K. Niazi, N. Mrachacz-Kersting, D. Farina, and K. Dremstrup, J. Neural Eng. **10**, (2013).

[11] R. Laje, T. Gardner, and G. B. Mindlin, Phys. Rev. E - Stat. Nonlinear, Soft Matter Phys. **64**, 056201/1 (2001).

[12] J. C. Lucero and L. L. Koenig, J. Acoust. Soc. Am. **117**, 1362 (2005).

[13] I. R. Titze, J. Acoust. Soc. Am. **83**, 1536 (1988).

[14] M. H. Krane, J. Acoust. Soc. Am. **118**, 410 (2005).

[15] B. H. Story and K. Bunton, J. Speech. Lang. Hear. Res. **53**, 1514 (2010).

[16] H. P. Zeigler and P. Marler, *Neuroscience of Birdsong* (Cambridge Univ Pr, 2008).

[17] M. Trevisan, G. Mindlin, and F. Goller, Phys. Rev. Lett. **96**, 1 (2006).

[18] E. M. Arneodo, Y. S. Perl, F. Goller, and G. B. Mindlin, PLoS Comput. Biol. **8**, e1002546 (2012).

[19] L. Goldstein, D. Byrd, and E. Saltzman, in *Action to Laguage via Mirror Neuron Syst.* (Cambridge University Press, 2006), pp. 215–249.

[20] J. W. Peirce, J. Neurosci. Methods **162**, 8 (2007).

[21] J. I. Hualde, A. Olarrea, and E. O'Rourke, *The Handbook of Hispanic Linguistics* (2012).

[22] A. M. Borzone de Manrique and M. I. Massone, J. Acoust. Soc. Am. **69**, 1145 (1981).

[23] P. Boersma and D. Weenink, (2013).

[24] M. F. Assaneo, M. A. Trevisan, and G. B. Mindlin, PLoS One **8**, e80373 (2013).

[25] F. Cuetos, M. Glez-Nosti, A. Barbón, and M. Brysbaert, Psicológica **32**, 133 (2011).

[26] M. F. Assaneo, J. I. Nichols, and M. A. Trevisan, PLoS One **6**, e28317 (2011).

[27] R. Kirchner, *An Effort Based Approach to Consonant Lenition* (Routledge, 2013).

[28] B. H. Story, J. Acoust. Soc. Am. **117**, 3231 (2005).

[29] B. H. Story and I. R. Titze, J. Phon. **26**, 223 (1998).

[30] M. M. Sondhi, J. Acoust. Soc. Am. **55**, 1070 (1974).

[31] J. Flanagan, J. Allen, and M. Hasegawa-Johnson, *Speech Analysis, Synthesis and Perception* (2008).

[32] B. H. Story, Physiologically-Based Speech Simulation Using an Enhanced Wave-Reflection Model of the Vocal Tract, University of Iowa, 1995.

[33] M. F. Assaneo, J. Sitt, G. Varoquaux, M. Sigman, L. Cohen, and M. A. Trevisan, Neuroimage **141**, 31 (2016).

[34] K. E. Bouchard, N. Mesgarani, K. Johnson, and E. F. Chang, Nature **495**, 327 (2013).

[35] D. F. Conant, K. E. Bouchard, M. K. Leonard, and E. F. Chang, J. Neurosci. 2382 (2018).

[36] M. Simonetta, M. Clanet, and O. Rascol, Electroencephalogr. Clin. Neurophysiol. Evoked Potentials **81**, 129 (1991).

[37] J. M. Correia, B. M. B. Jansma, and M. Bonte, J. Neurosci. **35**, 15015 (2015).

[38] T. J. Whitford, B. N. Jack, D. Pearson, O. Griffiths, D. Luque, A. W. F. Harris, K. M. Spencer, and M. E. Le Pelley, Elife 1 (2017).

Tables

TABLE I. ANOVA analysis with frequency of appearance of the CCVs in Spanish and trial number as independent factors. We also performed the tests using the absolute frequency values, with equivalent results.

|   |   | Frequency (low, medium, high) | Trial number |
|---|---|---|---|
| $\Delta$ |   | $p<10^{-5}$, $F=9.54$, $df=2$ | $p<10^{-5}$, $F=42.6$, $df=2$ |
| $\tau'$ | fricative-fricative | $p=0.004$, $F=8.37$, $df=2$ | $p=0.4$, $F=0.89$, $df=2$ |
|   | fricative-plosive | $p=6\times10^{-4}$, $F=7.51$, $df=2$ | $p=0.4$, $F=0.93$, $df=2$ |
|   | plosive-fricative | $p=0.002$, $F=9.93$, $df=1$ | $p=0.12$, $F=2.1$, $df=2$ |
|   | plosive-plosive | $p=0.62$, $F=0.24$, $df=1$ | $p=0.59$, $F=0.54$, $df=2$ |

Supplementary Materials

supplementary.zip contains the pool recorded of CCVs in wav format from our experiment, marked with the timing variables (in Praat format).